\documentclass{aa}
\usepackage{graphicx}
\usepackage{natbib}
\usepackage[usenames]{color}
\usepackage{threeparttable}

\newcommand{\SiIII}{\ion{Si}{iii}}

\newcommand{\NeVI}{Ne\ts$\scriptstyle{\rm VI}$}

\newcommand{\CaXIII}{Ca\ts$\scriptstyle{\rm XIII}$}
\newcommand{\CaXV}{Ca\ts$\scriptstyle{\rm XV}$}
\newcommand{\FeXIX}{Fe\ts$\scriptstyle{\rm XIX}$}

\def\ls{\mathrel{\lower4pt\vbox{\lineskip=0pt\baselineskip=0pt
           \hbox{$<$}\hbox{$\sim$}}}}
\def\gs{\mathrel{\lower4pt\vbox{\lineskip=0pt\baselineskip=0pt
           \hbox{$>$}\hbox{$\sim$}}}}
\def\drawbox#1#2{\hrule height#2pt

\hbox{\vrule width#2pt height#1pt \kern#1pt
              \vrule width#2pt}
              \hrule height#2pt}

\def\Asym#1#2{\vcenter{\vbox{\drawbox{#1}{#2}
              \kern-#2pt       
              \drawbox{#1}{#2}}}}

\usepackage{txfonts}
\begin{document}

\title{A nanoflare model for active region radiance: application of artificial
neural networks}

\author{M. Bazarghan\inst{1,2}
 \and H. Safari\inst{2,3,4}
\and D.E. Innes\inst{3} \and E. Karami\inst{5} \and S.K.
Solanki\inst{3}}

\institute{ IUCAA, Post Bag 4, Ganeshkhind, Pune 411 007, India
\and  Institute for Advanced Studies in Basic Sciences, Zanjan,
Iran \and Max-Planck Institut f\"{u}r Sonnensystemforschung,
 37191 Katlenburg-Lindau, Germany
\and Department of Physics, Zanjan University, Zanjan, Iran \and
Department of Electronics Science, University of Pune, Pune 411007, India}

\date{Received ----; accepted ----}

\abstract
{Nanoflares are small impulsive bursts of energy that blend
with and possibly make up much of the solar background emission.
Determining their frequency and energy input is central to understanding
the heating of the solar corona. One method is to extrapolate the
energy frequency distribution of larger individually observed
 flares to lower energies. Only if the power law exponent is
greater than 2 is it considered possible that nanoflares contribute significantly to the
energy input. }
{Time sequences of
ultraviolet line radiances observed in the corona of an active region are modelled with
the aim of determining
the power law exponent of the nanoflare energy distribution.}
{A simple nanoflare model based on three key parameters (the flare
rate, the flare duration, and the power law exponent of the flare
energy frequency distribution) is used to simulate emission line
radiances from the ions \FeXIX, \CaXIII, and \SiIII,
observed by SUMER in the corona of an active region as
it rotates around the east limb of the Sun. Light
curve pattern recognition by an Artificial Neural Network (ANN) scheme is
used to determine the values.
 }
{The power law exponents, $\alpha\approx2.8$, $2.8$, and 2.6 are obtained for
\FeXIX, \CaXIII, and \SiIII\ respectively. }
{The light curve simulations  imply a power law exponent greater than the critical
value of 2 for all ion species. This implies that
 if the energy of flare-like events
is extrapolated to low energies,  nanoflares could provide a significant
contribution to the heating of active region coronae.}


 \titlerunning{Estimating nanoflare distributions using ANN}

\maketitle

\section{introduction}
Heating the corona by the dissipation of current sheets
was first suggested by \citet{Gold64} and later developed to form the basis
of the
nanoflare heating model by \citet{Levine74} and \citet{Parker83, Parker88}.
The idea is that current sheets arise spontaneously in coronal
magnetic fields that are braided and twisted by random photospheric
footpoint motions. These current sheets dissipate
in many small-scale reconnection events,
heating  and accelerating plasma in the coronal loops.
In the corona, they would give rise to
multiple unresolvable loop strands with specific observable signatures
 \citep{Zirker94, WWH02, CK04, PK05}.  Recently \citet{Aschwanden08}
found evidence against such multi-temperature strands in TRACE coronal
images. He concludes that nanoflare heating is only possible if it occurs in
the chromosphere/transition region where heating across magnetic field lines can
 produce the isothermal loops seen in the corona.
Irrespective of where the nanoflare energy input sites are,
a key question is whether the energy of nanoflares is
sufficient to heat the corona or not.
Most of the individual nanoflares would be too small to detect and
the majority would be small fluctuations on the
overall background.
That background could be produced by the blending of many small events.

The approach taken to estimate their contribution has been
to extrapolate the energy frequency distribution of detectable flare-like events.
The energy frequency distribution of larger flares tends to follow a power law
distribution
\begin{equation}
{dN\over dE} \sim E^{-\alpha},
\end{equation}
where  $dN$ is the number of flares per energy interval $dE$.  The energy of
small flares dominates if  $\alpha>2$ \citep{Hudson91}. This is therefore
a critical parameter for the nanoflare heating model.
The standard
method to determine $\alpha$ is to evaluate the energy of many flares in
a series of observations and then plot their frequency in bins of energy $dE$.
 The majority of analyses based on this type of
event counting deduce $\alpha\approx 1.7$ \citep{Lin84, Shimizu95, AP02},
a value smaller than the critical 2. These results may, however,
be misleading.
For example, \citet{Parnell04} demonstrated that one can obtain
$\alpha$ ranging from 1.5 to 2.6 for the same data set using different but
still reasonable sets of assumptions for the analyses.

\begin{figure}
\includegraphics[width=8.1cm,angle=0]{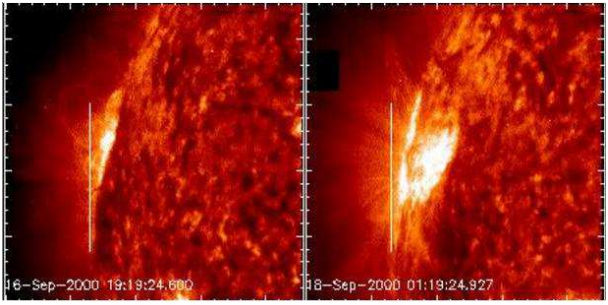}
 \caption{ EIT 195 \AA ~images of the observed active region at two times,
showing the position of the SUMER slit, indicated by the vertical line.}
\label{fig1}
\end{figure}

Here we take an alternative approach and model ultraviolet (UV) radiances
observed by the
Solar Ultraviolet Measurements of Emitted Radiation \citep[SUMER;][]{Wetal95, Wetal97}
in an active region corona, assuming
that the radiance fluctuations and the nearly constant `background'
emission are caused
by small-scale stochastic flaring \citep{PS04b, PS06}. The model
 has been applied successfully to UV radiance fluctuations in the quiet Sun
\citep{PS06}. The method compares light curves generated assuming random flaring
with a power law frequency distribution to
the light curves of an observed emission line.
It has the advantage that it takes into account without bias weak,
 blended micro- and nanoflares that
produce a nearly continuous background.

Here we apply this technique to off-limb time series recorded by SUMER.
The three lines modelled,
\FeXIX~$\lambda$\,1118.07 (6.3~MK), \CaXIII~$\lambda$\,1133.76 (2.2~MK) and
\SiIII~$\lambda$\,1113.23 (0.06~MK), cover two decades
of formation temperature from the lower transition region
to the hotter gas in the corona.

The analysis described here uses Artificial Neural Networks (ANNs) to
find the optimum match to the three parameters of the model.
The main advantage of this
method over previous analyses
based on the radiance distribution function   \citep{PS06, SISP07} is that
we are able to obtain quantitative values for all parameters,
including $\alpha$.
Another advantage of the ANN method is that it concentrates on the
number and shape of the emission peaks
along the light curves with little weight on the low radiance pixels,
which was a problem with the
\citet{SISP07} analysis.

\section{SUMER data and analysis }
\label{analysis}
\begin{figure}
\center
\includegraphics[width=8.0cm,height=4.8cm,angle=0]{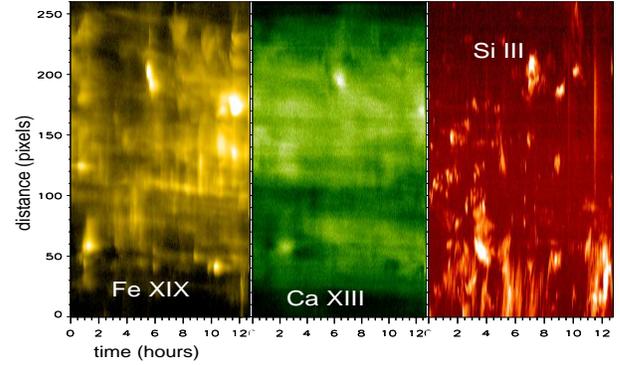}
 \caption{Time series of line radiance along the SUMER spectrometer slit for the
 period 16 Sep 19:00 UT  to 17 Sep 7:36 UT. The distance in pixels along the SUMER slit is
 shown on the vertical axis.}
 \label{fesica}
\end{figure}

The observed active region (AR 1967) is
shown in Fig.~\ref{fig1}. This is the region and data set
 discussed in
\citet{Wang06}.
The SUMER $300\arcsec\times4\arcsec$ slit was placed, as shown,
at a fixed position above the limb.
Observations with a cadence of 90~s
in six spectral lines,
\FeXIX\ $\lambda$\,1118.07 (6.3 MK), \CaXV\ $\lambda$\,1098.48 and
$\lambda$\,555.38 (3.5 MK), \CaXIII\ $\lambda$\,1133.76 (2.2 MK),
\NeVI\ $\lambda$\,558.62 (0.3~MK)
and \SiIII\
$\lambda$\,1113.23 (0.06 MK) were transmitted, for periods of
 12.6 hours followed by
a full spectrum scan (800$-$1600 \AA)  of 3.4 hours.
A typical time sequence in any one line  consists of
500 exposures.
The three strongest lines, \FeXIX, \CaXIII, and \SiIII, are analysed here.
Images of their radiance  along the slit
are shown  in Fig.~\ref{fesica} for a typical 12.6 hour period.
Distinct events can be seen in \FeXIX, but only the very strongest make
an impression on the bright active region \CaXIII\ emission when
they cool \citep{IW04b}.
\SiIII\ is seen
close to the limb and appears to be generated by small
surge-like ejections.
Our results are based on three such time series, taken over the days 16-18
September 2000.

The emission along each row was very noisy at several positions. To
improve the signal-to-noise but at the same time not to lose individual
structures, the light curves were obtained by first averaging
 SUMER data over five spatial pixels (5\arcsec) along the slit.
 Only the light curves with all 500 data points above a chosen threshold
 were selected for analysis. We did not want to base the
  threshold  on an absolute intensity because this would have biased the
  input data against low background. So for the
  \FeXIX\ and \CaXIII, the threshold was set such that the ratio of the \NeVI\ to
 \FeXIX\ intensity was less than 0.5. This ensures that only light curves
 from the central part of the active region were taken.
 Most of the  \SiIII\ emission was concentrated near
 the limb to the south (Fig.~\ref{fesica}). The \SiIII\ selection was based on the
 local scatter in the second moment of the line, the line width.
If the standard deviation of the line width was greater than 1.0 over a local
5x5 space-time block, then the central data point and associated
light curve were excluded from the analysis.
This resulted in 35 test light curves for both \FeXIX\ and \CaXIII\ and 11 test curves
for \SiIII.
Before being fed to the neural network,  all light curves were
normalized to their maximum.

\section{Model}
\subsection{Applicability}
The emission in the active region corona is assumed to be caused by many
random flares with flare radiances following a
power law frequency distribution.
Flares with a power law frequency distribution, $\alpha$, in
radiance are assumed to erupt with a frequency, $p_f$, and have a flare duration
$\tau = \tau_r + \tau_d$, where $\tau_r$
is the rise time and $\tau_d$ the decay  time. We assume $\tau_r/\tau_d = 0.5$.
The other free parameter in the model is the ratio of the maximum to  minimum
flare energy which is set to
    $E_{\rm max}/E_{\rm min}=50$.

\begin{figure}
\includegraphics[width=7.5 cm]{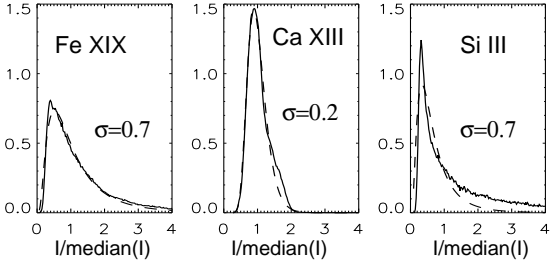}
 \caption{Distribution functions of
SUMER data in the active region corona (solid lines) and
best fit lognormal functions (dashed lines). The radiances are normalized to their
median and their distributions to the number of data points.}
\label{timeseries}
\end{figure}

For a large number of
independent random flares, the distribution of normalized radiances, $J = I/\overline{I}$
where $I$ is the radiance, is
lognormal with shape parameter $\sigma$ \citep{PS06}:
\begin{equation}\label{lognormal}
 f(J) = \frac{1}{\sigma J \sqrt{2 \pi}}exp \left(-\frac{(\log
 J)^2}{2\sigma^2}\right).
\end{equation}
$\sigma$ is inversely proportional to $\sqrt{\tau p_f}$ (Pauluhn
\& Solanki 2007), with a slight $\alpha$
dependence \citep{SISP07}.
A small shape parameter ($\sigma< 0.3$) indicates a symmetric
distribution due to high background emission caused by either a
long duration time, $\tau$, or a high flare frequency, $p_f$.
 The radiance distributions of the three lines of \FeXIX,~
\CaXIII~ and \SiIII~ and their lognormal fits are shown in Fig.~\ref{timeseries}.
This gives us confidence that the stochastic flare model is
applicable. It is interesting to note that both the \FeXIX\ and the \SiIII\ lines
have the same shape parameter.

\begin{figure}
\includegraphics[width=8.1 cm]{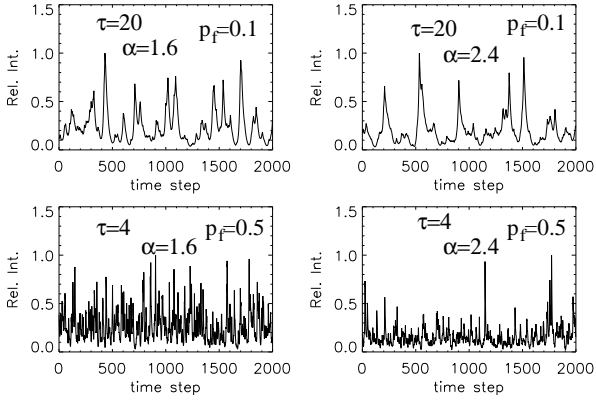}
 \caption{Light curves for flare models run with different $\tau$, $p_f$ and $\alpha$
 parameters. All light curves have $\tau p_f = 2$.}
\label{twoalphs}
\end{figure}

Light curves for the stochastic flare model are shown for $\alpha=1.6$ and $\alpha=2.4$,
and two combinations
of $\tau p_f=2$ in Fig.~\ref{twoalphs}.
The light curves are visibly different, although they all
 have shape parameter $\sigma \approx0.6$.
The effect of $\alpha$ on the light curve is seen
in the ratio of strong to weak flares. The left-hand light curves have more
large flares
because they have a smaller exponent. Picking up these pattern changes is
the strength of the ANN method.

\subsection{Neural networks and parameter estimation}
We applied
 the ANN method  to probe the unknown parameters (power law exponent, $\alpha$,
 duration time, $\tau$, and flare rate, $p_f$) of
 the three lines. ANNs have become a  popular tool in almost
every field of science. In recent years, ANNs have been widely
used in astronomy for applications such as star/galaxy
discrimination, \citep{Andreon00, Cortiglioni01}, morphological
classification of galaxies, \citep{Storrie92, Ball04}, and
spectral classification of stars \citep{Hippel94, SDSS08, ELODIE08}.

We employ probabilistic neural networks \citep[PNNs][]{Specht88,
Specht90}.
 The PNN learns to approximate the probability
density function of the training samples. It uses a supervised
training set to develop distribution functions within a pattern
layer. These functions in the recall mode are used to estimate
the likelihood of an input feature vector being part of a learned
category or class.

\begin{figure}[t]
\includegraphics[width=7.5 cm,angle=-0]{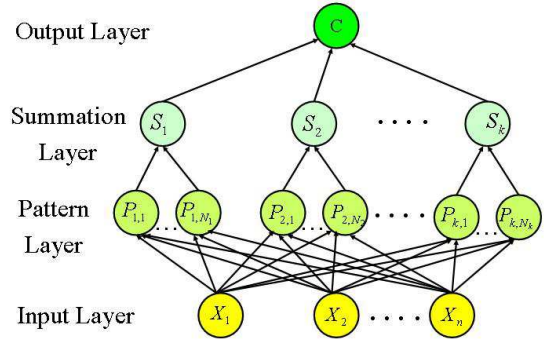}
\caption{Schematic of a typical probabilistic neural
network.} \label{PNN} \end{figure}

An example of a PNN
is shown in Fig.~\ref{PNN}. This network has four layers. The
network contains an input layer which has as many elements as
there are separable parameters needed to describe the objects to
be classified. It has a pattern layer, which organizes the
training set such that each input vector is represented by an
individual processing element.  The third
layer is the summation layer, which has as many processing
elements as there are classes to be recognized. Each element in
this layer combines via processing elements with the pattern layer
which relates to the same class and prepares that category for
output. Finally, there is the output layer that corresponds to the
summation unit with the maximum output.

For the identification of SUMER light curves, the input vector,
 $X = (x_1, x_2, ..., x_n )$, is the light curve with 500 data points
 ($n=500$).
The network is first trained to classify light curves corresponding
to all the possible combinations of $\alpha$, $\tau$, and $p_f$.
For this we synthetically generate light curves
with the nanoflare code described in \citet{PS06}.
We generate one light curve for each
combination of the parameters:
\begin{itemize}
     \item[-] the power law exponent spanning  $1.5\leq\alpha\leq3.2$ in
    steps of 0.1,
    \item[-] the duration time spanning  $1.5\leq\tau\leq52$ in steps of 1,
    \item[-] the flare rate spanning $0.1\leq p_f\leq0.9$ in steps of 0.1
    with additional values at $0.05$ and $0.95$.
\end{itemize}
This gives a  set of 6930 pattern groups ($k=6930$), one group for each combination
 of $\alpha$, $\tau$, and $p_f$.
Each pattern group, $k$, is characterized by $N_k$ Gaussian
functions \citep{Specht88, Specht90}.

When a SUMER light curve of an unknown classification is fed to the network, the
summation layer of the network computes the probability functions $S_k$ of
each class.
Finally at the output layer we have $C$, the value with the highest
probability.

\section{Results and conclusions}
\label{results}
\subsection{Results}

In the present work,  PNN is used
as a tool to extract the three flare model parameters required to
reproduce the SUMER light curves.  All 35 \FeXIX\ and \CaXIII, and 11
\SiIII\
SUMER light curves from
the three days of observations were fed individually into the neural
network and the parameters were obtained for each light curve separately.
The final PNN outputs are shown in Table
\ref{tab1}.
The bold numbers are the statistically maximum occurrence for each parameter.
For example for \FeXIX,  $\alpha=2.8$ is found in more than 70\% of the light
curves. The minimum and maximum values, given on the left and right, indicate
the scatter in the light curve parameters.

\begin{table}
  \centering
\caption{The SUMER spectral lines and the parameter values given by
PNN.}
\label{tab1}
\smallskip
\begin{threeparttable}
   \begin{tabular}{c c c c} \hline\hline\\
SUMER spectra  & \multicolumn{3}{c}{PNN outputs\tnote{1}}\\
lines & $\alpha$ & $\tau$ & $p_f$ \\
\hline \hline \\
\FeXIX & 2.5 {\bf 2.8} 3.1  & 9 {\bf 14} 20 & 0.1 {\bf 0.2} 0.7 \\
\CaXIII & 2.6 {\bf 2.8} 3.0 & 41 {\bf 45} 45 & 0.8 {\bf 0.9} 0.9 \\
\SiIII & 2.4 {\bf 2.6} 2.8 & 5 {\bf 9} 12  & 0.2 {\bf 0.3} 0.9 \\
 \hline
\end{tabular}
\begin{tablenotes}
\item[1]The most frequent values are given in bold, and the minimum and maximum
on
the left and right. $\tau$ is per exposure time (90 s) and $p_f$ is
per exposure time per
 5\arcsec$\times$4\arcsec\ spatial element.
\end{tablenotes}
\end{threeparttable}
\end{table}

In each line there is  20\% scatter in  $\alpha$, and 50\% scatter in
$\tau$.
The range of $p_f$ values for \FeXIX\ and \SiIII\ is much broader,
suggesting that events producing emission in these temperature ranges do not have the
same rate everywhere but are
seen in irregular bursts. We also note that the value of $\tau p_f$ is
roughly the same for both \FeXIX\ and \SiIII, as
suggested by their shape parameter (Fig.~\ref{timeseries}).
The \CaXIII\ light curves are all matched with a high
value of $p_f$, consistent with the idea  that the 1~MK active
region corona requires almost continuous flaring.
The four times higher rate for \CaXIII\ than \FeXIX\ suggests that most of the
\CaXIII\ emission is produced by heating events below the \FeXIX\ formation
temperature (6.6~MK).

\begin{figure}
\includegraphics[width=7.5 cm]{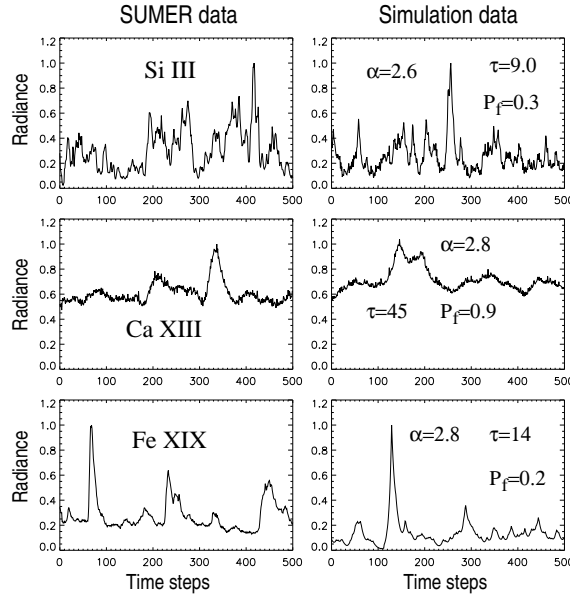}
\caption{Samples of the radiance time series: left
panel: SUMER data, and right panel: simulation data obtained with
the parameters given in Table 1. }
\label{lightcurves}
\end{figure}

Example light curves obtained using these parameters are compared with the observed
ones in Fig.~\ref{lightcurves}. Both the \SiIII\ and \CaXIII\ simulations
look remarkably similar to their observed light curves. The background radiance of
the \FeXIX\ light curve is about a factor of 2 too low. The \FeXIX\ light curves had
a $p_f$ ranging from 0.7 to 0.1, so we suspect that in this case the
 $p_f$ value is slightly too low. Also for \FeXIX,
 the ratio $\tau_r/\tau_d$ deduced from the data is smaller than
the fixed value 0.5 used here.
This may influence the accuracy of the method.

The sensitivity of the PNN output
depends on the training set.
 During the training session, the network must see all possible
patterns that it is supposed to classify in the testing session.
With 500 simulated light curves in the training set,
PNN was not able to converge for
several of the SUMER light curves.
When we increased the number
of simulated light curves to 6930, we were able to obtain unique parameters
 for all observed light curves.

\subsection{Conclusions}
The concept that the solar corona may
be heated by numerous, randomly distributed, small flare-like events called nanoflares is
considered by comparing simulated and observed emission line light curves.
The difference between this and previous methods is the fully automated
modelling of the light curve structure. There is no human decision required for
background/event cut-off levels or best fit parameters.

The result is power law flare energy frequency exponents greater than 2.5 for
all three emission lines considered, \SiIII, \CaXIII\ and \FeXIX.
This is consistent with the corona being heated
mainly by nanoflares, and
demonstrates the importance of nanoflare 'background' emission in determining the
power law exponents.
The parameter with highest uncertainty or largest scatter is the flare rate,
especially for the lines formed at transition region and hot flare temperatures.
Coronal plasma at these temperatures is produced sporadically and
is associated with more specific coronal and chromospheric loop structures than the
general active region corona, so the scatter is to be expected.

The next
step will be to determine the actual flare energies producing the
nanoflare emission. This is a much more complicated exercise
because the modelled light curves are observed in the corona which may be
heated by events occurring lower in the atmosphere \citep{Aschwanden08},
so that it requires a model for the energy transfer to the observation position.

\begin{acknowledgements}
 H. Safari
acknowledges the warm hospitality and financial support during his
research visit to the solar group, MPS.
\end{acknowledgements}

\bibliographystyle{aa}

\end{document}